\documentstyle[prb,aps,amsfonts,amssymb,amsmath,twocolumn,epsfig,bbm]{revtex}
\begin{document}
\draft
\sloppy

\newcommand{\w}{\omega} 
\newcommand{\kf}{k_F}
\newcommand{\zop}{{z_{\text{op}}}}
\newcommand{\zf}{{z_{F}}}
\newcommand{\kappaV}{{{\mbox{\boldmath $\kappa$}}}}
\newcommand{\surfint}{\int \!\!\!\! \int} 
\newcommand{\k}{{\vec{k}}} \newcommand{\q}{{ \vec{q}}}
\newcommand{\A}{\text{\AA}} 
\newcommand{\IM}{\text{Im}}
\renewcommand{\vec}[1]{{\bf #1}} 
\newcommand{\Q}{\vec{Q}}
\newcommand{\zz}{{\mathbbm{Z}}_2}

\wideabs{ 

\title{Some remarks about pseudo gap behavior
 of nearly antiferromagnetic metals}
  \author{A. Rosch} 
\address{Institut f\"ur Theorie der Kondensierten
    Materie, Universit\"at Karlsruhe, D-76128 Karlsruhe, Germany}
  \date{\today} 
\maketitle

\begin{abstract}
  In the antiferromagnetically ordered phase of a metal, gaps open on
  parts of the Fermi surface if the Fermi volume is sufficiently
  large.  We discuss simple qualitative and heuristic arguments under
  what conditions precursor effects, i.e.  pseudo gaps, are expected
  in the {\em paramagnetic} phase of a metal close to an
  antiferromagnetic quantum phase transition.  At least for weak
  interactions, we do not expect the formation of pseudo gaps in a
  three dimensional material.  According to our arguments, the upper
  critical dimension $d_c$ for the formation of pseudo gaps is
  $d_c=2$. However, at the present stage we cannot rule out a higher
  upper critical dimension, $2\le d_c \le 3$. We also discuss briefly
  the role of statistical interactions in pseudo gap phases.
\end{abstract}
}

Experiments on metals close to an antiferromagnetic quantum
critical point (QCP) show clearly that these systems cannot be
described by standard Fermi liquid theory. This is not very
surprising, as at the QCP magnetic fluctuations dominate and
electronic quasi particles scatter from spin-fluctuations
characterized by a diverging correlation length. Indeed, a theory of
quantum critical fluctuations interacting weakly with Fermi liquid
quasi particles \cite{hertz,millis} can explain a substantial part of
the experiments at least if effects like weak impurity scattering are
properly taken into account \cite{roschPRL}. However, a number of
experiments seems to contradict the standard spin-fluctuation
scenario, presently the best studied example for this is probably
CeCu$_{6-x}$Au$_x$ \cite{loehneysen,stockert,schroeder}. It has been
speculated that this might be due to anomalous two-dimensional spin
fluctuation\cite{stockert} or a partial breakdown of the Kondo
effect\cite{schroeder}.

In this paper we discuss a different route which can lead to a
breakdown of the theory of weakly interacting spin fluctuations, first
proposed by Hertz \cite{hertz,millis}. The general
idea\cite{schrieffer} is the following: close to the QCP, the behavior
of the system is dominated by large antiferromagnetic domains of size
$\xi$, slowly fluctuating on the time scale $\tau_\xi \sim \xi^\zop$
where $\zop$ is the dynamical critical exponent of the order
parameter. As $\xi$ is diverging when the QCP is approached, it is
suggestive to assume that the electrons will adjust their wave
functions adiabatically to the local antiferromagnetic background and
will therefore show a similar behavior as in the antiferromagnetically
ordered phase. If the Fermi surface is sufficiently large, the
(staggered) order parameter of the antiferromagnetic phase induces
gaps in parts of the Fermi surface with $\epsilon_{\vec{k}}\approx
\epsilon_{\vec{k}\pm \Q}\approx 0$, where $\epsilon_{\vec{k}}$ is the
dispersion of the quasi particles measured from the Fermi energy and
$\Q$ the ordering wave-vector of the antiferromagnet.  Will precursors of this
effect show up and induce pseudo gaps in the paramagnetic phase for
sufficiently large $\xi$? Pseudo gaps play an important role in the
physics of underdoped
cuprates\cite{andersonBook,kuebert,balents,franz,chubuOld,pseudogaps,abanov,tremblay}
and it has been speculated that they are indeed precursors of gaps in
either superconducting, antiferromagnetic, flux or striped phases. In
this report we want to investigate qualitatively on the basis of
simple physical arguments under what generic conditions such pseudo
gaps are expected to occur close to an antiferromagnetic QCP. We will
consider only systems, where the ordered antiferromagnet is metallic, therefore
our discussion might have less relevance for the high T$_c$
superconductors where the undoped antiferromagnet is a (Mott-) insulator.

\begin{figure}
\begin{center}
\epsfig{width=0.4 \linewidth,file=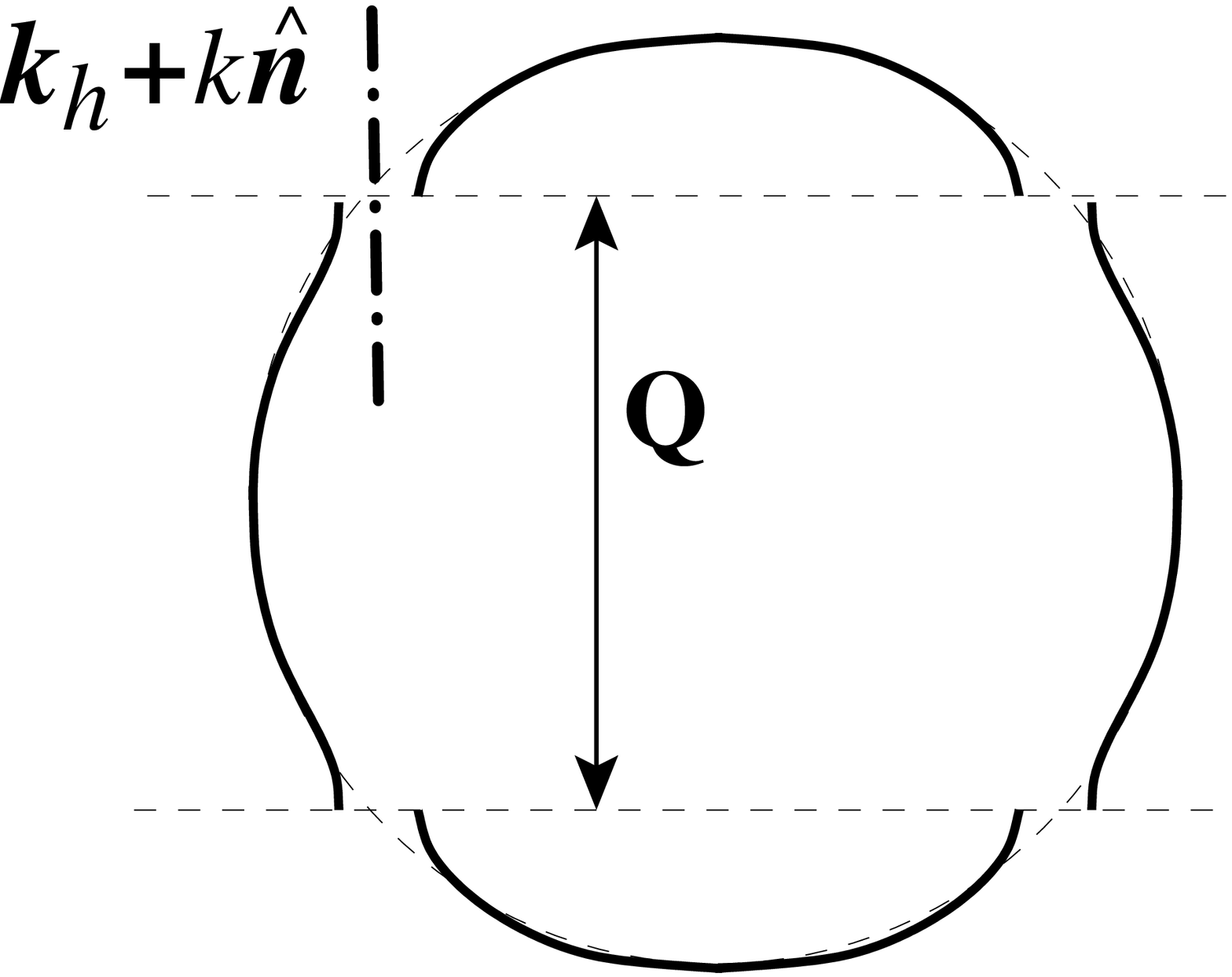}
\end{center}
\caption{Schematic plot of the Fermi surface. In the ordered phase of
  a metallic antiferromagnet gaps open at the boundaries of the
  magnetic Brilloin zone.
\label{fermiFig}
}
\end{figure}
To define the concept of a pseudo gap more precisely, we first analyze
the ordered phase where in mean field theory the Hamilton of the
electrons is of the form
\begin{eqnarray}
H_\Delta&=&\sum_{\sigma,\vec{k}} 
(c^\dagger_{\sigma, \vec{k}},c^\dagger_{\sigma, \vec{k}+\Q}) 
\left(
\begin{array}{cc}
\epsilon_{\vec{k}} & \sigma \Delta \\
\sigma \Delta & \epsilon_{\vec{k}+\Q}
\end{array}
\right)
\left(
\begin{array}{c}
c_{\sigma, \vec{k}} \\
c_{\sigma, \vec{k}+\Q}
\end{array}
\right). \label{Hmf}
\end{eqnarray}
$\Delta$ is proportional to the staggered order parameter
(assumed to point in $z$ direction) and the $\vec{k}$ sum extends over
a magnetic Brillouin zone. Close to the ``hot lines'' on the Fermi surface
 (``hot points''
in two dimensions) with $\epsilon_{\vec{k}_h}=
\epsilon_{\vec{k}_h\pm \Q}=0$, a gap opens (see Fig.~\ref{fermiFig})
and the band structure at $\vec{k}=\vec{k}_h+\delta \vec{k}$ is
approximately given by 
\begin{eqnarray}
 \epsilon^{\pm}_{\vec{\delta k}}\approx \frac{1}{2}
\left((\vec{v}_1+\vec{v}_2) \delta \vec{k} \pm
  \sqrt{\left((\vec{v}_1-\vec{v}_2) \delta \vec{k}\right)^2 +4
    \Delta^2}\right) \label{epsMF}
\end{eqnarray}
where $\vec{v}_1=\vec{v}_{\vec{k}_h}$ and
$\vec{v}_2=\vec{v}_{\vec{k}_h+\Q}$ are the Fermi velocities close to
the hot points. The gap is e.g. visible if one integrates the spectral
function $A_{\vec{k}}(\w)$ for $\vec{k}$-vectors along a direction
$\hat{n}$ in the $(\vec{v}_1, \vec{v}_2)$ plane perpendicular to
$\vec{v}_1+\vec{v}_2$ (dash-dotted line in Fig.~\ref{fermiFig})
$\tilde{A}(\w)=\int dk A_{\vec{k}_h+k \hat{\vec{n}}}(\w)$.  In mean
field theory $\tilde{A}(\w)$ displays a well-defined gap of size $2
\Delta$. This gap is a consequence of the reduced translational
symmetry and is expected to be present in the ordered phase of the
antiferromagnet, even in a regime, where the predictions of mean field
theory are quantitatively wrong. Interactions of quasi particles far
away from $\vec{k}_h$ with each other and with the spin-fluctuations
will actually induce some small weight within these (renormalized)
gaps but this does not invalidate the mean field picture:
$\tilde{A}(\w)$ vanishes rapidly in the limit $\w \to 0$ in the
ordered phase as it is obvious from the usual Fermi liquid phase space
arguments.  From general scaling arguments one expects in the {\em
  paramagnetic} phase close to the QCP, that at $T=0$
\begin{eqnarray}\label{pseudoF}
\tilde{A}(\w) \sim \w^{\alpha} f(1/(\w \xi^{z}))
\end{eqnarray} 
with $f(x\to 0)\approx const.$ and $f(x\to \infty)\sim x^\alpha$ where
$z$ is a dynamical critical exponent (see below).  Within mean field
theory no precursor of the gaps show up and $\alpha=0$.  However, one
would expect $\alpha>0$ if the wave function of the quasi particles
adjusts adiabatically to the local antiferromagnetic order on length
scales smaller than $\xi$.

\begin{figure}
\begin{center}
\epsfig{width=0.95 \linewidth,file=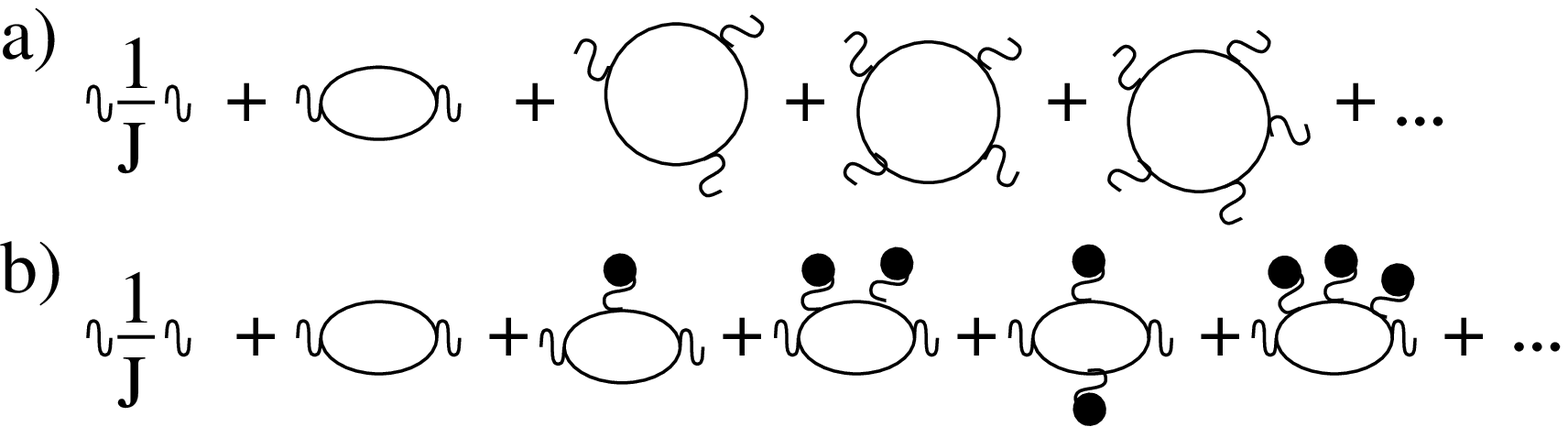}
\end{center}
\caption{a)
  Effective action $S[\Phi]$ according to Hertz\cite{hertz} for the
  model defined in Eqn.~(\ref{Sall}) 
 after the electrons have been integrated out.
  The lines denote free Greens functions $G_0(x-x',\tau-\tau')$ of the
  electrons, the wiggles are the fields $\Phi(x,\tau)$.  b) Quadratic
  part of the effective action $\Phi \to \langle \Phi \rangle+\delta
  \Phi$ in the ordered phase ( {\protect
    \epsfig{width=0.23cm,file=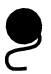}} denotes the order
  parameter $\langle \Phi \rangle$).  Combinatorial prefactors are
  omitted both in a) and b).
\label{actionFig} }
\end{figure}
This paper focuses on the $T=0$ behavior directly at the QCP as we are
mainly interested in the question whether pseudogaps affect the
quantum critical behavior and therefore  (\ref{pseudoF}) with $\alpha>0$
serves as a defintion for a pseudogap.  Note that pseudogap physics can be
considereably more pronounced in other regimes, e.g. for nearly
magnetic metals with Heisenberg or xy symmetry in $d=2$ for low, but
finite temperatures in a parameter regime, where the system is deep in
the ordered phase at $T=0$. This regime has for example been
investigated in detail by Vlik, Tremblay {\it et al.} \cite{tremblay}.

For definiteness, we will consider a model of Fermions  $f_{\vec{k}\sigma}$
coupled linearly to a collective bosonic field  $\Phi_{\vec{q}}$
with the following action in imaginary time\cite{hertz}
\begin{eqnarray}
S&=&\int_0^\beta\!\!\!  d \tau  \Biggl[
\sum_{\sigma,\vec{k}} 
f^{*}_{\sigma,\vec{k}} (\partial_\tau+\epsilon_{\vec{k}}) 
f_{\sigma,\vec{k}} \nonumber+ \sum_{\vec{q}}
 \vec{\Phi}_{\vec{q}}^* \frac{1}{J_{\vec{q}}}  
\vec{\Phi}_{\vec{q}} \nonumber \\
&& \qquad + \sum_{\vec{k}\vec{q}i\alpha\beta} 
 \Phi^i_{\vec{q}}  f_{\alpha, \vec{k}+\vec{q}}^\dagger \sigma^i_{\alpha \beta} 
 f_{\beta, \vec{k}}+h.c.\Biggr]\label{Sall}
\end{eqnarray}
where $\sigma^i$ are the Pauli matrices and $\beta=1/T$ 
 the inverse temperature. Integrating out the
collective field induces a spin-spin interaction $J$ of the Fermions.
For realistic models one should also add charge-charge interactions,
which are, however, not expected to change the physics close to a
magnetic QCP qualitatively.

 Many years ago, Hertz\cite{hertz} has proposed to describe the QC
metallic antiferromagnet in the spirit of a Ginzburg-Landau-Wilson approach
in terms of  a fluctuating order parameter $\Phi(x,\tau)$ with an effective
action
\begin{eqnarray}
S&=&S_0+S_{\text{int}} \label{S}\\
S_0&=&\frac{1}{\beta} \sum_{\vec{k} ,\w_n} \Phi_{\vec{k}\w_n}^* (r+
(\vec{k}\pm\Q)^2+\gamma |\w_n|) 
\Phi_{\vec{k}\w_n} \label{phi4}\label{S0}\\
S_{\text{int}}&=& U \int_0^\beta d \tau \int d^d \vec{r} |\Phi(x,\tau)|^4 
\end{eqnarray}
where $\w_n= 2 \pi n/\beta$
are bosonic Matsubara frequencies and $\Phi_{\vec{k}\w_n}$ is the
Fourier transform of $\Phi(x,\tau)$. The term linear in $\w_n$ is due
to the scattering from quasi particles which induce the Landau damping
of the spin-fluctuations. As discussed in detail in the original paper
by Hertz \cite{hertz}, the action $S$ describes only the leading terms
in an expansion which is derived by integrating out the Fermions in
(\ref{Sall}). 
The expansion is shown  schematically in Fig.~\ref{actionFig}a
(due to time-reversal symmetry, cubic terms vanish in the limit
$\w_n\to 0$).  A simple scaling analysis\cite{hertz} with $k\sim 1/L$,
$\w \sim 1/L^\zop$, $\Phi(x,\tau)\sim L^{1-(d+\zop)/2}$ with $\zop=2$
shows that the interaction term $S_{\text{int}}$ vanishes $\sim
1/L^{d+\zop-4}$, i.e.  $U$ is (dangerously) irrelevant in dimensions
$d>4-\zop=2$.  Furthermore, higher order interactions and frequency
and momentum dependencies of the effective vertices are even more
irrelevant. A pseudogap as it is defined in (\ref{pseudoF}) would
certainly change the critical exponent $\zop$, as it would strongly
reduce the damping of the spin-fluctuations.  As the scaling analysis
sketched above does give no indications for such a phenomenon for
$d\ge 4-\zop$, it strongly suggests that a strong-coupling effect like a
pseudo gap should never occur in dimensions $d>4-\zop$ at least as long as
the (bare) interactions are not too strong.

This line of arguments (which would be completely valid close to a
classical phase transitions) is {\em not} reliable in the case of a
quantum phase transition in a metal. This can be seen for example by
considering the {\em ordered} phase. An expansion of the Hertz action
(\ref{S}) around the mean field $\Phi=\langle \Phi \rangle + \delta
\Phi$ suggests that the transverse spin-fluctuations (assuming
Heisenberg or xy symmetry) are damped.  However, the Goldstone theorem
guarantees that the spin-waves are not damped in the limit $\w,\vec{k}\to
0$. The physical origin of this is essentially the same as in the
previous discussion of pseudo-gap formation: the wave function of the
electrons adjust to the slowly varying antiferromagnetic background.  A simple RPA
approximation based upon the mean field Hamiltonian (\ref{Hmf})
correctly describes this effect on a qualitative level. It is
therefore instructive to investigate how the RPA contribution arise in
the effective action $S[\Phi]$. In Fig.~\ref{actionFig}b it is shown
that spin-spin interactions $\Phi^n$ of arbitraryly high order $n$ are
needed to recover the trivial RPA+mean-field result.

Two scenarios seem to be possible to resolve the apparent conflict
that contributions which are irrelevant by power counting are
important in the ordered phase.  The first possibility is that all the
higher interactions are indeed {\em irrelevant} in the sense that the
physics of the formation of undamped spin fluctuations does not
influence the quantum critical behavior on the paramagnetic side of
the phase diagram in any qualitative manner -- in technical terms,
they are ``dangerously irrelevant'' and important only in the ordered
phase. The analysis given below suggests that this situation is
actually realized in three dimensions.  The second possibility is that
pseudo gap formation is important and that spin-spin interactions of
arbitrarily high order have to be kept which implies that (\ref{S})
does not describe the physics properly and the ``true'' critical
theory cannot be formulated in terms of the order parameter alone but
has to include fermionic modes. For example, the spin fluctuation
theory of the cuprates as it is worked out by Abanov and Chubukov
\cite{abanov} suggests such a scenario in $d=2$. What can
go wrong with the simple scaling arguments given above? Belitz,
Kirkpatrick {\it et al.}\cite{belitz} have recently shown in their
analysis of the dirty nearly ferromagnetic metal that scaling is
indeed not reliable due to a very simple physical reason: The Hertz
action implies that a domain of size $\xi$ fluctuates very slowly on
the time scale $\tau_\xi \propto \xi^{\zop}$ with $\zop=2$.  However,
in a clean metal there is a much faster and more efficient way to
propagate information from one side of a fluctuating domain to the
other: ballistic electrons can traverse the domain in the time
$\tau_{F} \propto \xi^{\zf}$ with $\zf=1$.  This defines a second
dynamical critical exponent $\zf$ (which can be renormalized due to
scattering from spin fluctuations, see below).  Power-counting is not
reliable because {\em two} different dynamical exponents $\zop$ and
$\zf$ exist simultaneously -- while there is only one large length
scale $\xi$, two rather different time scales exist. The question
which of these scales is relevant for a given process generally
requires a detailed analysis and is not at all obvious. This physics
should therefore be investigated in a careful renormalization group
calculations which includes both fermionic and bosonic degrees of
freedom. We will not try such an analysis here but instead use a
properly modified scaling argument to investigate the possibility of
pseudo gap formation.

For our scaling analysis\cite{d2}, 
we assume that the susceptibility at the QCP
is of the form suggested by (\ref{S0}) ($d\ge 2$) 
\begin{eqnarray}
\chi_{\vec{q}\pm \Q}(\w)\sim \frac{1}{\vec{q}^2+(i \w)^{2/\zop}}.\label{chi}
\end{eqnarray}
We are mainly interested in the case $\zop=2$, smaller values for
$\zop$ might be relevant if pseudo gap formation takes place
\cite{schrieffer}, larger values have e.g. been used to fit
experiments\cite{schroeder} in CeCu$_{6-x}$Au$_x$ and have been
claimed\cite{abanov,chubuPrivate} to be relevant in $d=2$. It is not
difficult to generalize the following arguments for susceptibilities
with other $\vec{q}$ and $\w$ dependencies\cite{chubuPrivate}.

The strategy of the following scaling analysis is to estimate the
effective amplitude of the quasistatic collective field seen by the
electrons. Obviously the answer will depend on which time- and
length-scale the electrons probe the background magnetization. The
main idea is, that a lower bound for the relevant time- and length
scales can be derived from Heisenberg's uncertainity relation and the
effective size of the gap.  The main assumptions of the following
arguments are discussed in detail in the second half of the paper:
we assume that above the upper critical dimension
for pseudogap formation, the nature of the electrons is not changed
completely by the quantum critical fluctuations.  According to the
mean field result (\ref{epsMF}) a gap of size $\w^*=\Delta$ opens in a
($d-2$ dimensional) stripe in momentum space of width
$k^*=\Delta/v_F$. Below, we will discuss the effect of interactions
which can change this relation to $\w^* \sim (k^*)^\zf \sim
\Delta^\zf$ where $\zf=1$ is the mean field exponent.  Heisenberg's
uncertainty relation dictates that the electrons have to see a
quasi-static antiferromagnetic background for a time $\tau^* \gtrsim
1/\w^*$ on a length scale of order $\xi^* \gtrsim 1/k^*$ perpendicular
to the direction of the hot lines to develop the pseudo gap. What is
the effective size of the quasi-static antiferromagnetic order
$\langle \Phi \rangle^{\text{eff}}_{\xi^*,\tau^*}$ on these length and
time scales?  The following estimate should at least give an upper
bound at the QCP
\begin{eqnarray}
\left(\langle \Phi \rangle^{\text{eff}}_{\xi^*,\tau^*}\right)^2 &\lesssim& 
\int_0^{\w^*} \!\!\!\!\!\! d\w  \!\! \int_{q_\perp < {k^*}}  
\!\!\!\!\!\!\!\!\! d^2 q_{\perp}  \!\! 
\int_{\infty}^{\infty}  \!\!\!
d^{d-2}q_{\|} \,\, \IM \chi_{\vec{q}\pm\Q}(\w) \label{phiEff}  \\
&\sim& (k^*)^{d+\zop-2}+(k^*)^2 (w^*)^{\frac{d+\zop-4}{\zop}} 
\nonumber\\
&\sim& \Delta^{(d+\zop-4)\frac{\zf}{\zop}+2}\label{phiEffRes}
\end{eqnarray}
where the anisotropic integration of $q$ takes into account that the
momentum of the electrons {\em parallel} to the hot line can vary on the
scale $k_F$. In (\ref{phiEffRes}) we assumed $\zf \le \zop$.  For our
scaling argument, it does not matter whether we use $\IM \chi(\w)$ or
e.g. $\chi(-i \w)$ in (\ref{phiEff}), the version given above is
motivated by the estimate of the quasi-elastic weight
obtained in a $T=0$ neutron scattering experiment with limited
resolution $\w^*$ and $k^*$.

If we assume furthermore that $\Delta$ is proportional to $\langle
\Phi \rangle^{\text{eff}}_{\xi^*,\tau^*}$ as suggested by the mean
field analysis (which should be valid above the upper
critical dimension), we obtain the inequality $\Delta^2 \lesssim
const.\cdot \Delta^{(d+\zop-4)\frac{\zf}{\zop}+2}$. This implies that,
at least in a weak coupling situation, pseudo gaps should appear only
if
\begin{eqnarray}
d+\zop \le 4 \label{pseudoCrit}
\end{eqnarray}
which is the central result of this paper. We believe, that it is
accidental that (\ref{pseudoCrit}) coincides with the condition for
the relevance of the $\Phi^{4}$ interaction (\ref{phi4}) in the Hertz
model as is evident from the fact that $z_F$ enters the inequality
(\ref{phiEffRes}). Within the approach of Hertz, $\zop=2$ and the
critical dimension for pseudo gap formation is therefore $d_c=2$. From
our scaling arguments we cannot say much about what will happen in
$d=d_c=2$ (or for $d<d_c$).  Based on the observation, that the
ordered phase is not well described by (\ref{S}), we suspect that the
Hertz description of a quantum critical antiferromagnet is {\em not}
valid in $d=2$ -- this point of view agrees with the results of Abanov
and Chubukov \cite{abanov} who have analyzed the spin-fermion problem
in $d=2$ in a certain large $N$ expansion.  In the pseudo gap phase we
expect by comparison to the ordered phase that $1 \le \zop < 2$.
Therefore it seems to be possible that the critical dimension is not
two but somewhere between 2 and 3 (Abanov and Chubukov claim
\cite{abanov}, however, that $\zop$ is larger than 2 in $d=2$
depending on the number of hot spots).  In three dimensions,
pseudo-gap formation will probably not invalidate the Hertz approach,
at least for weak coupling.

The derivation of (\ref{pseudoCrit}) is far from being
rigorous and based on a number of assumptions. In the following two of
them, which are probably the most important ones,
 are discussed in more detail. First we consider non-Fermi liquid
effects due to the scattering from singular spin fluctuations, the
second aspect concerns strong coupling effect and the respective role
of amplitude and angular fluctuations of the staggered magnetization.

The scattering from spin fluctuations strongly modifies the
quasi-particles close to the hot lines. In leading order perturbation theory,
the self energy of those electrons at $T=0$ is given by
\begin{eqnarray}
\IM \Sigma_{\vec{k}}(\Omega)\approx  g_S^2 \sum_{\vec{k}'}
\int_0^{\Omega}\! \!\!  d \w \IM
\chi_{\vec{k}-\vec{k}'}(\w) \IM g^0_{\vec{k'}}(\w-\Omega), \label{sigma}
\end{eqnarray}
where $g_S$ is the vertex of the coupling of electrons to
spin-fluctuations (here, we assume the absence of pseudo gap
formation and therefore $g_S$ is finite) and $g^0_{\vec{k'}}(\w)\approx
1/(\w-\epsilon_{\vec{k}}+i 0^+)$ is the Greens function of the (free) fermions.
Using (\ref{chi}) and (\ref{sigma}) we obtain at the QCP
\begin{eqnarray}
\IM \Sigma_{\vec{k}_h+\delta \vec{k}}(\Omega)\sim \Omega^{1+\frac{d-3}{\zop}}
f\!\left(\frac{(\delta \kappa)^2}{\Omega^{2/\zop}}\right) 
\end{eqnarray}
where $\delta \kappa \sim \delta \vec{k} \cdot \vec{v}_{\vec{k}_h+\Q}$
is a measure for the distance from the hot line and $f$ is some
scaling function with $f(x\to 0) \sim const.$ and $f(x\to \infty)\sim
1/x^{\frac{5-d}{2}}$.  For $\zop=2$ and far away from the hot lines,
Fermi liquid behavior is recovered.  Our previous arguments suggest
that typical frequencies and momenta for the pseudo gap formation are
$\delta \kappa \sim \Delta$ and $\Omega\sim \Delta^{\zf}$, therefore
the typical argument $\Delta^{2 (1-\zf/\zop)}$ of $f$ is small and the
momentum dependence of $\IM \Sigma$ can be neglected for $\zf<\zop$
and will not induce new effects for $\zf=\zop$ (this is the reason,
why we used $k^* \sim \Delta$ in our scaling analysis).  From this we
obtain $\Sigma(\Omega_{\text{typical}})\sim
\Omega^{1+\frac{d-3}{\zop}}$.  Below three dimensions, the quasi
particle picture breaks down close to the hot lines and therefore some
of our perturbative arguments might fail\cite{abanov}.  Ignoring this
possibility, we conclude that typical energies $E_{\vec{k}}$ of the
(incoherent) fermionic excitations are determined from $E_{\vec{k}}+c
E_{\vec{k}}^{1+\frac{d-3}{\zop}} \sim v_F (k-k_F)$ (because we can
neglect the $\vec{k}$ dependence of $\Sigma$) and therefore
\begin{eqnarray}
\zf=\max\!\left[1,\frac{\zop}{d+\zop-3} \right]
\end{eqnarray}
which is the value which should be used in our previous arguments for
$d+\zop \ge 4$, i.e. in the absence of pseudo gap formation.  An
effect which we haven't taken into account in our discussion, is that
generically, close to the antiferromagnetic QCP, a superconducting
phase is stabilized \cite{super}, however, at least in $d\ge 3$ the
ordering temperature of the superconductor $T_c$ is usually much
smaller than the typical scale $T^*$ below which the quantum critical
behavior of the antiferromagnet dominates. In $d=2$ the situation might be
different \cite{super,schmalian} with $T_c \sim c T^*$, where $c$ is
a constant of order 1.

It is important to emphasize, that our estimate (\ref{phiEff}) of
$\langle \Phi \rangle^{\text{eff}}$ and therefore our main result
(\ref{pseudoCrit}) is based on the assumption that {\em amplitude}
fluctuations of the staggered order parameter are present and can be
described by (\ref{chi}).  Electrons adjust their wave functions much
better to angular fluctuations of the direction of the staggered
magnetization than to fluctuations of its size, because a rotation of
the spin-quantization axis does not cost any energy in the long wave
length limit (assuming weak spin-orbit coupling and/or a sufficient
high symmetry of the underlying crystal). This adiabatic adjustment is
not included in our estimates. Numerical results of Bartosch,
Kopietz\cite{bartosch}, Millis and Monien\cite{monien} show that in
$d=1$ amplitude and phase fluctuations have a drastically different
effect on pseudo gap formation.  Nevertheless, our approach to focus
on amplitude fluctuations in our previous discussion was valid as
within the theory of Hertz (\ref{S}), the interactions of
spin-fluctuations are irrelevant and amplitude fluctuations exist for
$d>2$. If they are present they should be the dominating mechanism to
destroy pseudo gap behavior. Below the upper critical dimensions, one
expects that amplitude fluctuations are frozen out and only angular
fluctuations dominate the critical regime.  Even in dimensions larger
than $2$ such a picture might be appropriate in a strong coupling
regime, e.g. if one considers a Heisenberg model with a large
antiferromagnetic coupling $J_{AF}$ coupled to a metal.
Unfortunately, the behavior of electrons in such a situation is much
less understood.  To investigate the pseudo gap phase in this case,
probably the most obvious theoretical route\cite{trans,schrieffer} to
describe the adiabatic adjustment of the wave function of the
electrons is to rotate the quantization axis of the electrons into the
{\em local} direction of the slowly fluctuating order parameter. This
approach has been used by a number of authors interested in the pseudo
gap phase of the cuprates\cite{kuebert,balents,franz}.  A natural
model to discuss this type of physics consists of a non-linear
$\sigma$-model coupled to the spin $\vec{S}(\vec{r})=\frac{1}{2}
f^{\dagger}_\alpha(\vec{r}) \boldsymbol{\sigma}_{\alpha,\beta}
f_\beta(\vec{r}) $ of fermions $f$.  The non-linear $\sigma$-model
$S_\sigma$ describes the directional fluctuations of the staggered
order parameter $\vec{n}$ in the absence of amplitude fluctuations.
The action in terms of $\vec{n}$ with $\vec{n}^2=1$ and the Grassmann
fields $f$ is given by\cite{shraiman}
\begin{eqnarray}
S&=&S_f+S_{\sigma}+S_{f\sigma} \\
S_f&=& \int_0^\beta\!\!\!  d \tau  \sum_{\sigma,\vec{k}} 
f^{*}_{\sigma,\vec{k}} (\partial_\tau+\epsilon_{\vec{k}}) f_{\sigma,\vec{k}} \nonumber \\
S_\sigma&= &\frac{1}{g}\int_0^\beta \!\!\!d \tau \int  \!\! d^d \vec{r}  (\partial_\tau \vec{n})^2+
(v \partial_\vec{r} \vec{n})^2 \nonumber\\
S_{f\sigma}&=& \Delta  \int_0^\beta\!\!\!  d \tau \int \!\! d^d \vec{r}  \cos(\Q \vec{r})\, 
\vec{n}(\vec{r},\tau)  \vec{S}(\vec{r},\tau). \nonumber
\end{eqnarray}
We have not written down the proper spin Berry phase which is
essential to describe the Kondo lattice correctly.  For simplicity, we
focus in the following on a model with an $O(2)$ symmetry
$\vec{n}=(0,\sin \phi(\vec{r},t),\cos \phi(\vec{r},t))$ and comment
below on the more difficult situation with $O(3)$ symmetry.  To
describe the pseudo gap, we define new fields $c$ with a quantization
axis rotated in the local direction of the order 
parameter\cite{trans,schrieffer}.
 \begin{eqnarray}
 \left(\begin{array}{c} c_\uparrow(\vec{r},\tau) \\ c_\downarrow(\vec{r},\tau)
   \end{array} \right)=\exp\!\left[i \Phi(\vec{r},\tau) 
\frac{\sigma^x}{2} \right]
\left(\begin{array}{c} f_\uparrow(\vec{r},\tau) \\ f_\downarrow(\vec{r},\tau) 
   \end{array} \right). \label{gauge}
\end{eqnarray}
The new fields $c$, which we call ``pseudo fermions'' in the
following, do not transform under a global rotation around $x$-axis,
this implies a separation of spin and charge degrees of freedom
\cite{balents,franz} if the low energy excitations are well described
by $c$ (see below).  The advantage of the transformation is, that
$S_{f\sigma}$ now describes the scattering of the pseudo fermions from
a {\em static} order parameter pointing always in $z$ direction which
can be treated non-perturbatively. The pseudo fermions are the natural
degrees of freedom in a situation, where the the single-particle wave
function adjusts to the (collective) magnetic background. If one
neglects the residual interactions with $\vec{n}$, gaps open along the
hot lines and the action of the pseudo fermions is given by
\begin{eqnarray}
S_{c}=\int_0^{\beta}(\sum_{\sigma,\vec{k}}
c^{*}_{\sigma \vec{k}} \partial_\tau c_{\sigma \vec{k}} +
H_{\Delta}(c^*,c))
\end{eqnarray}
 where $H_{\Delta}(c^\dagger,c)$ is the mean field
Hamiltonian (\ref{Hmf}).

The residual interaction of $\vec{n}$ and $c$ arises from the Berry
phase $f^{*}\partial_\tau f$ and kinetic energy of the
electrons. The semi-classical contributions $S^{\text{sc}}_{c\sigma}$
is given by the minimal substitution which corresponds to the gauge
transformation (\ref{gauge}).  Using $(\partial_\mu \phi)
\frac{\sigma^x}{2}=((\partial_\mu \vec{n})\times \vec{n}) \cdot
\frac{\boldsymbol{\sigma}}{2}$, we obtain
\begin{eqnarray} \label{semi}
S^{\text{sc}}_{c\sigma}= - i \int_0^{\beta} \!\!\!\! d\tau \! \int_{-\infty}^{\infty} 
\!\!\!\! d^d\vec{r}\,
(\left[\partial_\tau+(\vec{v}_F \boldsymbol{\nabla}))\vec{n}\right]\times \vec{n}) \cdot \vec{S}.
\end{eqnarray}
In the notation used here, the Fermi velocity $\vec{v}_F$ is
actually a function of $\vec{k}=-i \nabla$ which acts on
$c_{\sigma}(\vec{r})$ hidden in $\vec{S}(\vec{r})$. The action
$S^{\text{sc}}_{c\sigma}$  describes an interaction of
spin-currents.

As the vertex in (\ref{semi}) vanishes in the limit $\w,\vec{k}\to 0$,
Schrieffer\cite{schrieffer} has argued that the effect of
$S^{\text{sc}}_{c\sigma}$ is small close to the QCP and that therefore
pseudo gaps and the associated decoupling of spin-fluctuations from
the fermions are a generic property of an antiferromagnetic QCP. This
argument is however misleading. One reason is that amplitude
fluctuations will destroy the pseudo gap in many relevant situations
as discussed above.  But even in the absence of amplitude
fluctuations, the pseudo fermions interact strongly with the magnetic
fluctuations by a pure quantum effect which is not included in the
semi-classical $S^{\text{sc}}_{c\sigma}$.  Formally, the origin of the
effect is that the rotation of a Fermion by $2 \pi$ changes its sign!
If $\phi(\vec{r},\tau)$ in (\ref{gauge}) jumps from $2 \pi$ to $0$,
the pseudo fermion $c$ abruptly flips its sign, giving rise to a huge
contribution to the effective action. There are many
possibilities\cite{franz,balents} to keep track of these sign changes
in a path integral, one of them is to rewrite the problem as a local
$\zz$-gauge theory where the arbitrary sign $\pm 1$ is the origin of
the $\zz$ symmetry. Here we follow a slightly different route by
replacing $\phi$ in (\ref{gauge}) by $\tilde{\phi}$ with
\begin{eqnarray}\label{phiTilde}
\tilde{\phi}(\tau,\vec{r})&=& 
\phi(0,0)+\int_{(0,0)}^{(\tau,\vec{r})}  (\partial_\mu \phi) d r^\mu = 
\phi(\tau,\vec{r})+ 2 \pi n.
\end{eqnarray}
The line integral is along some path in space-time, e.g. $r^\mu(u)= (
u \tau, u \vec{r})$, where $u$ varies in the interval $[0,1]$ and
$\mu=0,1,..,d$ denotes the temporal and spatial
directions\cite{derivative}.  The integer $n$ in (\ref{phiTilde}) is
defined in such a way that $\tilde{\phi}$ is continuous along the path
$r^\mu(u)$ and therefore the pseudo fermions, defined by replacing
$\phi$ by $\tilde{\phi}$ in (\ref{gauge}), will vary smoothly without
sudden sign changes along $r^\mu(u)$. But along some other paths,
abrupt sign changes are unavoidable.  This is obvious by considering
the line integral $\int_{(\tau,\vec{r})}^{(\tau,\vec{r})}
(\partial_\mu \phi) d r^\mu= 2 \pi n$ along some {\em closed} path in
space time. By definition it has to be a multiple of $2 \pi$ and $n$
is obviously the number of magnetic vortices of the xy-model enclosed
in the loop. From this we conclude that the pseudo fermions acquire a
phase $\pi$, i.e. a minus sign, whenever they circle around a magnetic
vertex: this is nothing but the well known Berry phase of a spin
forced to move on a circle.  A possible interpretation of this result
is, that each xy- vortex has attached to its core a magnetic flux with
half a flux quantum. The pseudo
electrons are  strongly interacting with the fluctuating magnetic
 vortices of the antiferromagnet and it is not obvious whether the gap
will survive. A likely possibility is that the interactions are so
strong that they are leading to confinement at least in some parameter
regime as it has been suggested in the context of the $\zz$ gauge
theory of fluctuating superconductors \cite{balents}.  One possible
way of confinement is the binding of the pseudo fermions to the
magnetic excitations in such a way, that the resulting degree of
freedom is nothing but the original electron $f_{\sigma \vec{k}}$. In
this case, we do not expect any pseudo gaps.  As we are not aware of
methods which can describe such a confinement transition, it is
difficult to give an estimate under which conditions a pseudo gaps
will occur in the model (\ref{S}). We can only speculate that the
formation of pseudo gaps might be controlled by the area-density of
vortices, i.e. the number of vortices per area $n_A$ piercing through
a given area in space time at the QCP, to be compared to
$(\Delta/v_F)^2$. Both $n_A$ and $\Delta$ are non-critical at the
transition.  If these are the relevant parameters, then pseudo gap
behavior is expected only if the density of vortices at the QCP is
small.

If the magnet has O(3) instead of xy symmetry, one can follow the same
steps which have been discussed before and one faces again the
problem, that statistical interactions are induced as soon as pseudo
fermions are introduced.  K\"ubert and Muramatsu\cite{kuebert} have
proposed in the context of a theory of a slightly doped t-J model a
convenient way to keep track of this statistical interaction with the
help of a CP$^1$ representation of $\vec{n}$ using two complex fields
$z_1$ and $z_2$ with $|z_1|^2+|z_2|^2=1$ and $\vec{n}=z^*_\alpha
{\boldsymbol{\sigma}}_{\alpha \beta} z_\beta$. In this language the
pseudo fermions interact strongly with the CP$^1$ fields via a local
$U(1)$ gauge theory\cite{kuebert}. Again, confinement seems possible.

In this paper, we have investigated the possibility of pseudo gap
behavior close to the QCP of a nearly antiferromagnetic metal. Based
on heuristic scaling arguments we suggest that generically, amplitude
fluctuations destroy pseudo gaps in dimensions $d>2$. In three
dimensions we expect that the Hertz theory is valid at least for not
too strong coupling while in $d=2$ it is probably modified due to
pseudo gap formation and the strong interaction of spin-fluctuations
and fermionic modes. These questions should be studied in a
renormalization group treatment of both fermionic and bosonic modes.
We were not able to derive any criteria for pseudo gap formation in a
situation where amplitude fluctuations are completely frozen out and
emphasized that the motion of the fermions on top of the spin
background leads to strong statistical interactions of the fermionic
modes with the excitations of the magnet.

I would like to thank A.V. Chubukov, F. Evers, M.~Garst, A. Mirlin, A.
Muramatsu, J.  Schmalian, A.-M. S. Tremblay and P.  W\"olfle for
helpful discussions and the Emmy Noether program of the DFG for
financial support.


\begin{references}
\bibitem{d2} The ``hot lines'' defined by $\epsilon_{\vec{k}_h}=
  \epsilon_{\vec{k}_h+\vec{Q}}=0$ are $d-2$ dimensional, therefore our
  scaling analysis applies only for $d \ge 2$.
\bibitem{derivative} The derivative $\partial_\mu \Phi=((\partial_\mu
  \vec{n})\times \vec{n})\cdot(1,0,0)$ is defined on a circle, i.e. it is smooth when $\phi$ jumps from $2 \pi$ to $0$.
\end{references}
\end{document}